\documentstyle[epsfig]{jfm}

\catcode`\œ=\active \gdefœ{\setbox0=\hbox{0}\hbox to\wd0{}}
\def\dynpercm{\nobreak\mbox{$\;$dynes\,cm$^{-1}$}}

\def\Real{\mbox{Re}}      
\def\Imag{\mbox{Im}}      
\def\Rey{\mbox{\it Re}}   
\def\Pran{\mbox{\it Pr}}  
\def\Pen{\mbox{\it Pe}}   
\def\Ai{\mbox{\rm Ai}}    
\def\Bi{\mbox{\rm Bi}}    
\ifCUPmtlplainloaded
\else
  \def\upi{\pi} 
  \def\umu{\mu} 
  \def\upartial{\partial} 
  \def\BbbE{\mbox{\sf E}} 
\fi
\ifCUPmtlplainloaded
  \def\BbbE{\Bbb E}
  
  \renewcommand{\simeq}{\approx}
\fi
\def\ssC{\mbox{\sf C}}         
\def\sfsP{\mbox{\sfs P}}       
\def\slsQ{\mbox{\sls Q}}       
\ifCUPmtlplainloaded
  \font\sfs = mtssi10 at 10.5pt   
  \font\sls = mtssbi10 at 10.5pt  
  \font\bit = mtmib10 at 10.5pt \skewchar\bit ='177  
\else
  \font\sfs = cmssi10  
  \font\sls = cmssi10  
  \font\bit = cmmib10 \skewchar\bit ='177  
\fi
\ifCUPmtlplainloaded
\else
  \font\tenbmi=cmmib10 at 10pt  \skewchar\tenbmi ='177
  \font\sevenbmi=cmmib10 at 7pt \skewchar\sevenbmi ='177
  \font\fivebmi=cmmib10 at 5pt  \skewchar\fivebmi ='177

  \newfam\bmifam
  \textfont\bmifam=\tenbmi
  \scriptfont\bmifam=\sevenbmi
  \scriptscriptfont\bmifam=\fivebmi
  
\fi
\newsavebox{\thalfbox}
\sbox{\thalfbox}{$\textstyle\frac{1}{2}$}
\newcommand{\thalf}{\usebox{\thalfbox}}

\newsavebox{\shalfbox}
\sbox{\shalfbox}{$\scriptstyle\frac{1}{2}$}
\newcommand{\shalf}{\usebox{\shalfbox}}

\newsavebox{\squartbox}
\sbox{\squartbox}{$\frac{1}{4}$} 
\newcommand{\squart}{\usebox{\squartbox}}

\newsavebox{\etbox}
\sbox{\etbox}{\boldmath$\eta$}
\newcommand{\etb}{\usebox{\etbox}}

\newcommand{\biS}{\mbox{\boldmath $S$}}

\newsavebox{\astrutbox}
\sbox{\astrutbox}{\rule[-5pt]{0pt}{20pt}}
\newcommand{\astrut}{\usebox{\astrutbox}}

\ifnfsstwo

\fi
\ifnfssone
  \newmathalphabet{\mathit}
    \addtoversion{normal}{\mathit}{cmr}{m}{it}
    \addtoversion{bold}{\mathit}{cmr}{bx}{it}

\fi
\ifoldfss

\fi

\newcommand{\GaPQ}  {G_a(P,Q)}
\newcommand{\Gat}   {\widetilde{G_a}}
\newcommand{\GsPQ}  {G_s(P,Q)}
\newcommand{\kgd}   {k\gamma d}
\newcommand{\ndq}   {\frac{\mbox{$\partial$}}{\mbox{$\partial$} n_q}}
\newcommand{\p}     {\mbox{$\partial$}}

\newcommand{\sumjm} {\sum_{j=1}^{M}}
\newcommand{\tti}   {\rightarrow\infty}
\newcommand{\ttz}   {\rightarrow 0}
\mathchardef\varLambda="0103

\newcommand{\ra}{\right\rangle}
\newcommand{\la}{\left\langle}

\ifCUPmtlplainloaded
  \newcommand{\pvi}{\int_0^{\infty}\mskip -30mu -\quad} 
\else
  \newcommand{\pvi}{\int_0^{\infty}\mskip -33mu -\quad} 
\fi
\ifCUPmtlplainloaded
  \let\bcdot=\undefined
  \NewSymbolFont{bldsym}{mtbsy10}{'60}
  \NewMathSymbol{\bcdot}{2}{bldsym}{01}
\else
  \font\tenbms=cmbsy10          \skewchar\tenbms ='60
  \font\sevenbms=cmbsy10 at 7pt \skewchar\sevenbms ='60
  \font\fivebms=cmbsy10 at 5pt  \skewchar\fivebms ='60

  \newfam\bmsfam
  \textfont\bmsfam=\tenbms
  \scriptfont\bmsfam=\sevenbms
  \scriptscriptfont\bmsfam=\fivebms

  \edef\bsy{\hexnumber\bmsfam}
  \mathchardef\bnabla="0\bsy72
  \mathchardef\bcdotsymbol="0\bsy01
  \def\bcdot{\,\bcdotsymbol\,}
\fi
\def\eg{{e.g.\ }}
\def\etc{{etc.\ }}
\def\etal{\mbox{\it et al.\ }}

\title[Maximum Drag Reduction]{Maximum Drag Reduction Asymptotes and the Cross-Over to the Newtonian
Plug}

\author[R. Benzi, E. De Angelis, I. Procaccia, V.S. L'vov, V. Tiberkevich]{R. BENZI$^{1}$, E. DE ANGELIS$^2$
 V. S. L'VOV$^3$, I.
PROCACCIA$^{3}$ and V. TIBERKEVICH$^3$}
\affiliation{$^1$ Dip. di Fisica and INFM, Universit\`a ``Tor
Vergata", Via della Ricerca Scientifica 1, I-00133 Roma, Italy,
$^2$ Dip. Mecc. Aeron., Universit\`a di Roma ``La Sapienza",
Via Eudossiana 18, 00184, Roma, Italy,  
$^3$ Dept. of Chemical Physics, The Weizmann Institute of Science,
Rehovot, 76100 Israel.}

\pubyear{2002}
\volume{538}
\pagerange{1--10}
\date{13 may 2004 and in revised form ??}
\newcommand {\bce}  {\begin{center}}
\newcommand {\ece}  {\end{center}}
\newcommand {\be}   {\begin{equation}}
\newcommand {\ba}   {\begin{array}}
\newcommand {\bea}  {\begin{eqnarray}}
\newcommand {\bfi}  {\begin{figure}}
\newcommand {\ee}   {\end{equation}} 
\newcommand {\ea}   {\end{array}}
\newcommand {\eea}  {\end{eqnarray}}
\newcommand {\efi}  {\end{figure}}

\newcommand {\VK}   {von-K\'{a}rm\'{a}n }

\def\Rset {{\rm I \kern-.2em R}} 
\def\mathbbH {{\rm I \kern-.2em H}} 
\def\mathbbC {{\rm I \kern-.6em C}} 

\begin{document}
\maketitle
\begin{abstract}
We employ the full FENE-P model of hydrodynamics of a dilute
polymer solutions to derive a theoretical approach to drag reduction in wall
bounded turbulence. We recapture the results of a recent
simplified theory which derived the universal Maximum Drag
Reduction (MDR) asymptote, and complement that theory with a
discussion of the cross-over from the MDR to the Newtonian plug
when the drag reduction saturates. The FENE-P model gives rise to
a rather complex theory due to the interaction of the velocity
field with the polymeric conformation tensor, making analytic
estimates quite taxing. To overcome this we develop the theory in
a computer-assisted manner, checking at each point the analytic
estimates by Direct Numerical Simulations (DNS) of viscoelastic
turbulence in a channel.
\end{abstract}

\section{Introduction}
\label{section1} The onset of turbulence in fluid flows is
accompanied by a significant increase in the drag
\cite{69Lu,00SW}. This drag poses a real technological hindrance
to the transport of fluids and to navigation of ships. It is
interesting therefore that the addition of long chain polymers
to wall-bounded turbulent flows can result in a significant
reduction in the drag. The basic experimental knowledge of the
phenomenon had been reviewed and systematized by 
\cite{75Virk}; the amount of drag depends on the characteristics
of the polymer and its concentration, but cannot exceed a
universal asymptote known as the ``Maximum Drag Reduction" (MDR)
curve which is independent of the polymer's concentration or its
characteristics. When the concentration is not large enough, the
mean velocity profile as a function of the distance from the wall
follows the MDR for a while and then crosses back to a
Newtonian-like profile, cf. Fig. \ref{profiles} left.

\bfi[t!] 
          \centerline{ 
          \hspace{0.3cm}
          \epsfig{figure=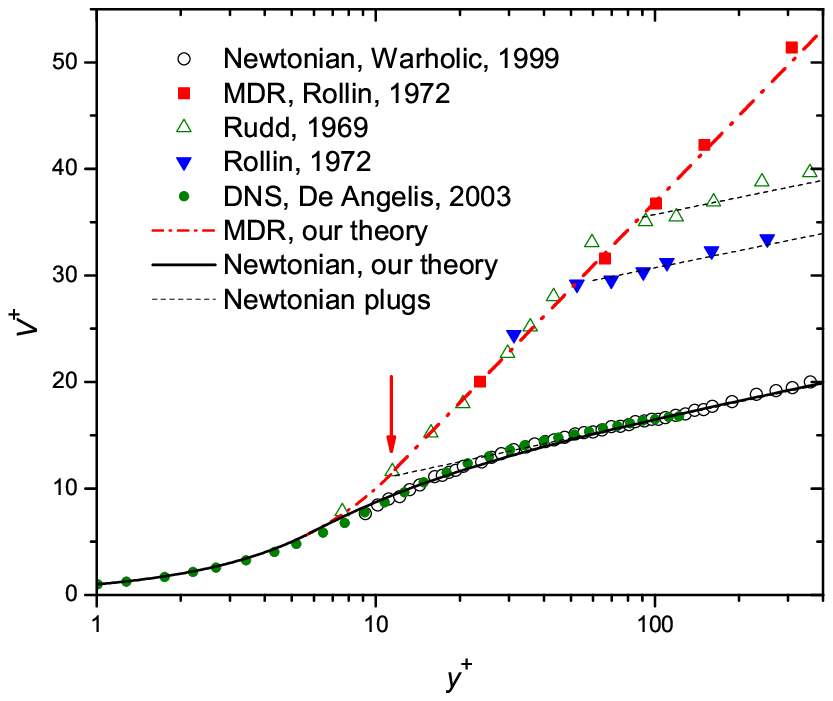,height=5.5cm} 
          \hspace{-0.4cm}
          \epsfig{figure=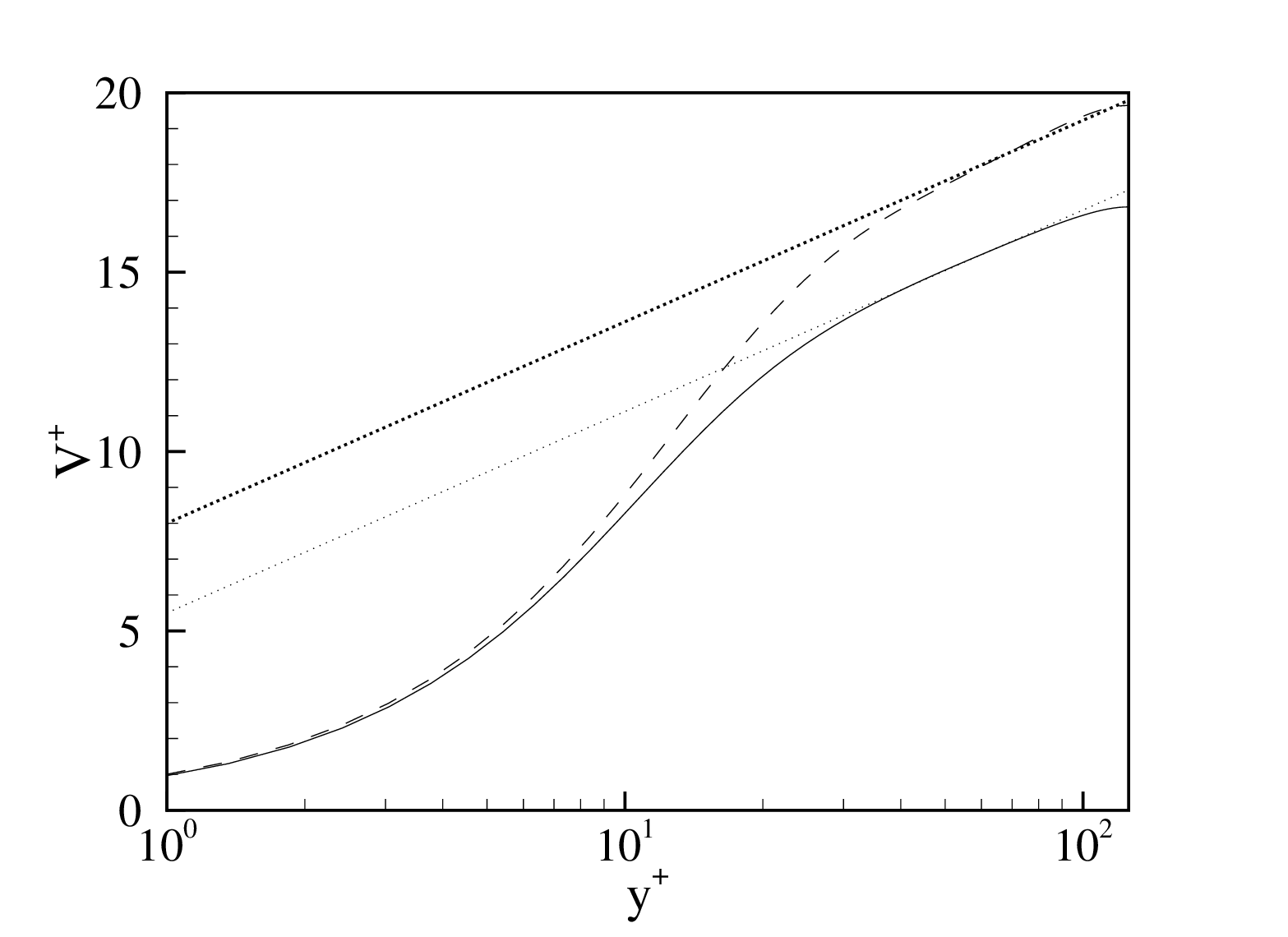,height=5.9cm} 
           } 
\caption{Left panel:Mean normalized  velocity profiles as a function of the
normalized  distance from the wall during drag reduction. The data
points from numerical simulations (green circles) 
and the experimental points (open circles) 
represent
the Newtonian results.  
The red data points (squares) 
represent the Maximum Drag Reduction (MDR) asymptote. The dashed
red curve represents the theory discussed in the paper 
which agrees with
the universal MDR. The arrow marks the crossover from the viscous
layer to the  Newtonian log-law of the wall. The blue filled
triangles 
and green open triangles 
represent the crossover, for intermediate concentrations of the
polymer,  from the MDR asymptote to the Newtonian plug.
Right panel: Mean velocity profiles for the Newtonian and for the
viscoelastic simulations with ${\rm Re}_\tau=125$.  Solid line:
Newtonian. Dashed line: Viscoelastic. The straight lines represent
the classical {\VK} log-law. The coordinate $y^+$ is the distance from
the wall in Pradtl'units $y^+ = v_*L/\nu$, where $v_*^2 = p'L$.
\label{profiles} }
\efi

Recently the nature of the MDR and the mechanism leading to its
establishment were rationalized, using a phenomenological theory
in which the role of the polymer conformation tensor was modeled
by an effective viscosity \cite{03LPPT}. It turned out that the
effective viscosity attains a self-consistent profile, increasing
linearly with the distance from the wall. With this profile the
reduction in the momentum flux from the bulk to the wall
overwhelms the increased dissipation that results from the
increased viscosity. Thus the mean momentum increases in the bulk,
and this is how drag reduction is realized. In \cite{04DCLPPT} it
was demonstrated by DNS that Navier-Stokes flows with viscosity
profiles that vary linearly with the distance from the wall indeed
show drag reduction in close correspondence with the phenomenon
seen in full viscoelastic simulations. The aim of this paper is to
complement the simplified theory with a derivation of the same
results on the basis of a full viscoelastic model of the
hydrodynamics of dilute polymer solutions. Such a derivation
forsakes some of the simplicity of the phenomenological theory,
but on the other hands it clarifies the role of the polymer
conformation tensor in furnishing viscosity-like contributions.
The relative complexity of the statistical theory is handled
by employing 
suitable approximations for the leading terms and  
by
computer assisted estimates of competing terms, allowing us at the
end to provide a compact theory with sharp predictions. In
addition to lending further support to the model of
\cite{03LPPT}, we will offer a discussion of the non-universal
saturation of drag reduction as a function of concentration,
length of polymer and relaxation time in various conditions of
experimental interests.

In Sect. 2 we consider the FENE-P model of viscoelastic flows and
review shortly the evidence for drag reduction in DNS of this
model. In Sect. 3 we employ the FENE-P model to derive a
statistical  theory of drag reduction in wall bounded
turbulence. In Sect. 4 we use the theory to predict the cross-over
from the universal MDR to the Newtonian plug when the conditions
differ from those necessary for attaining the MDR. In Sect. 5 we
present a summary and a discussion.

\section{Equations of motion for viscoelastic flows and drag reduction}
\label{section2} 
Viscoelastic flows are represented well by hydrodynamic equations
in which the effect of the polymer enters in the form of a
``conformation tensor" $ R_{ij}(r,t)$ which stems from the
ensemble average of the dyadic product of the end-to-end distance
of the polymer chains \cite{87BCAH,94BE}. A successful model that
had been employed frequently in numerical simulations of turbulent
channel flows is the FENE-P model \cite{87BCAH}. Flexibility and finite
extensibility of the polymer chains are reflected by the
relaxation time $\tau$ and the Peterlin function $P(r,t)$ which
appear in the equation of motion for $R_{ij}$:
\begin{eqnarray}
\frac{\partial R_{\alpha\beta}}{\partial t}+( u_{\gamma}  \nabla_{\gamma})
R_{\alpha\beta}
&&=\frac{\partial u_\alpha}{\partial r_\gamma}R_{\gamma\beta}
+R_{\alpha\gamma}\frac{\partial u_\beta}{\partial r_\gamma}\nonumber\\
&&-\frac{1}{\tau}\left[ P(r,t) R_{\alpha\beta} -\rho_0^2
\delta_{\alpha\beta} \right]\ ,\label{EqR}\\
P(r,t)&&=(\rho_m^2-\rho_0^2)/(\rho_m^2 -R_{\gamma\gamma}) \ .
\end{eqnarray}
In these equations $\rho^2_m$ and $\rho^2_0$ refer to the maximal
and the equilibrium values of the trace $R_{\gamma\gamma}$. Since
in most applications $\rho_m\gg \rho_0$ the Peterlin function can
also be written approximately as $P( r,t)\approx (1/(1 -\alpha
R_{\gamma\gamma})$ where $\alpha=\rho_m^{-2}$. In its turn the
conformation tensor appears in the equations for fluid velocity
$ u_{\alpha}( r,t)$ as an additional stress tensor:
\begin{eqnarray}
&&\frac{\partial  u_{\alpha}}{\partial t}+( u_{\gamma}\nabla_{\gamma}) u_{\alpha}=-\nabla_{\alpha} p
+\nu_s \nabla^2  u_{\alpha} + \nabla_{\gamma} {T_{\alpha\gamma}}+ F_{\alpha}\ , \label{Equ}\\
&&{T_{\alpha\beta}}( r,t) = \frac{\nu_p}{\tau}\left[\frac{P(
r,t)}{\rho_0^2}  R_{\alpha\beta}(r,t) - \delta_{\alpha\beta} \right] \ . \label{T}
\end{eqnarray}
Here $\nu_s$ is the viscosity of the neat fluid, $F_{\alpha}$ is the
forcing and $\nu_p$ is a viscosity parameter which is related
to the concentration of the polymer, i.e. $\nu_p/\nu_s\sim
c_p$ where $c_p$ is the volume fraction of the polymer. We
note however that the tensor field can be rescaled to get rid of
the parameter $\alpha$ in the Peterlin function, $\tilde
R_{\alpha\beta}=\alpha R_{\alpha\beta}$ with the only consequence
of rescaling the parameter $\nu_p$ accordingly. Thus the actual
value of the concentration is open to calibration against the
experimental data. Also, in most numerical simulations, the term
$P\rho_0^{-2} R$ is much larger than unity and the unity tensors in (\ref{EqR}) and (\ref{T}). 
Therefore, in the theoretical development below we shall use
the approximation
\begin{equation}
	T_{\alpha\beta} \sim \frac{\nu_p}{\tau}\frac{P}{\rho_0^2} R_{\alpha\beta}.
\label{approx}
\end{equation}
These equations were simulated on the computer in a channel or
pipe geometry, reproducing the phenomenon of drag reduction in
experiments. The most basic characteristic of the phenomenon is
the increase of fluid throughput in the channel for the same
pressure head, compared to the Newtonian flow. This phenomenon is
demonstrated in Fig. \ref{profiles} right taken from \cite{03ACLPP}. 
As one can see the simulation
is limited compared to experiments; the Reynolds number is
relatively low, and the MDR is not attained. Nevertheless the
phenomenon is there, and we will be able to use the simulation to
asses and support the theoretical steps taken in the theoretical
development. We should note that in the simulations one needs to
add a small artificial viscosity to Eq. (\ref{EqR}), in the form
of a term $\kappa\Delta R_{\alpha\beta}$. This is done solely for
taming numerical instabilities, and has very little consequence on
the theory presented below, where it is not taken into account.

\section{The derivation of the MDR}
\label{wallbounded}
\subsection{Exact balance equations}
Consider the fluid velocity $ U_{\alpha}(r)$ as a sum of
its (time) average and a fluctuating part:
\begin{equation}
U_{\alpha}(r,t) =  V_{\alpha}(y) +  u_{\alpha}( r,t) \ ,  V_{\alpha}(y)
\equiv \langle  U_{\alpha}( r,t) \rangle \ . \label{split}
\end{equation}
For a channel of large length and width all the averages, and in
particular $ V_{\alpha}(y) \rightarrow V(y)\delta_{\alpha y}$, are functions of $y$ only.
The objects that enter the theory are the mean shear $S(y)$, the
Reynolds stress $W(y)$, the kinetic energy $K(y)$ and the mean
conformation tensor $\langle R(y)\rangle$; the first three are
defined respectively as
$$
S(y)\equiv d V(y)/d y \ , \;\; W (y)\equiv - \langle u_x u_y\rangle \ ,
\;\; K(y) = \langle |u|^2\rangle/2 \ .
$$
Taking the long time average of Eq. (\ref{Equ}), and integrating
the resulting equation along the $y$ coordinate produces and exact
equation for the momentum balance:
\begin{equation}
W + \nu S + \frac{\nu_p}{\tau} \langle P R_{xy}\rangle (y) =
p'(L-y) \ . \label{1}
\end{equation}
where $p'$ is the pressure gradient and $L$ is the mid-channel height. 
The RHS is simply the rate at which momentum is produced by the
pressure head, and on the LHS we have the Reynolds stress which is
the momentum flux, the viscous dissipation of momentum, and the
rate at which momentum is transferred to the polymers. Near the
wall it is permissible to neglect the term $p'y$ on the RHS for
$y\ll L$. Next we want to derive the balance equation for energy.
To this aim we need to take into account the extra energy
dissipation due to transfer of energy from the velocity field to
the polymers. This energy dissipation, denoted as $\epsilon_p$, in the limit of validity of 
(\ref{approx}) can be exactly computed to be
\begin{equation}
\epsilon_p = \frac{\nu_p}{2\tau^2}\langle P^2 {\rm Tr}
R_{\alpha\beta}\rangle \ , \label{epsp}
\end{equation}
where Tr$R_{\alpha\beta}\equiv R_{xx}+R_{yy}+R_{zz}$. As stated in Eq.
(\ref{split}), we split the velocity field into its mean and the
fluctuation, and consider separately the balance equation for the
mean energy $V^2$ and for the turbulent energy $\langle
u^2\rangle$. The former yields an equation identical to (\ref{1})
but multiplied by $S$:
\begin{equation}
WS + \nu S^2 + \frac{\nu_p}{\tau} \langle P R_{xy}\rangle (y)S
= p'LS \ . \label{balmean}
\end{equation}
On the other hand the balance equation for the turbulent energy
cannot be written exactly. To understand how to write it, we need
to identify first the meaning of the third term on the LHS of Eq.
(\ref{balmean}). To gain insight on how to represent properly this
term we consider how to estimate correlations containing the
tensor $R_{ij}$. Using equation (\ref{EqR}), we obtain on the
average:
\begin{equation}
\frac{\partial}{\partial y} \la u_y R_{ij} \ra =
-\frac{1}{\tau}\langle P
R_{ij}\rangle+ \la R_{ik} \partial_k u_j \ra +
\la R_{jk} \partial_k u_i \ra \ .
\label{3}
\end{equation}

\subsection{Closure approximations}
To proceed, we employ a 1-point closure approximation, keeping
only terms that survive this closure. In this approximation, the
only terms $\langle\partial_ku_jR_{ik}\rangle$ which survive are
proportional to $S$, i.e.
\begin{equation}
\langle R_{ik}\partial_k u_j\rangle = a_i S \langle R_{iy}\rangle \delta_{jx}
\label{1point}
\end{equation}
where $a_i$ are constant of order $1$. For $i=x,y$, using
(\ref{1point}), equations (\ref{3}) give:
\begin{eqnarray}
\label{xx}
\frac{1}{\tau} \langle P\rangle \langle R_{xx}\rangle &=& 2 a_x \langle
R_{xy}\rangle S\ , \label{xx}        \\
\frac{1}{\tau} \langle P\rangle \langle R_{xy}\rangle &=& a_y \langle
R_{yy}\rangle S \ . \label{xy}
\end{eqnarray}
To test the quality of this closure we check each term by DNS,
comparing between the exact and approximated forms.
A description of the DNS simulations is reported in \cite{03ACLPP}.

\bfi[t!] 
          \centerline{ 
          \epsfig{figure=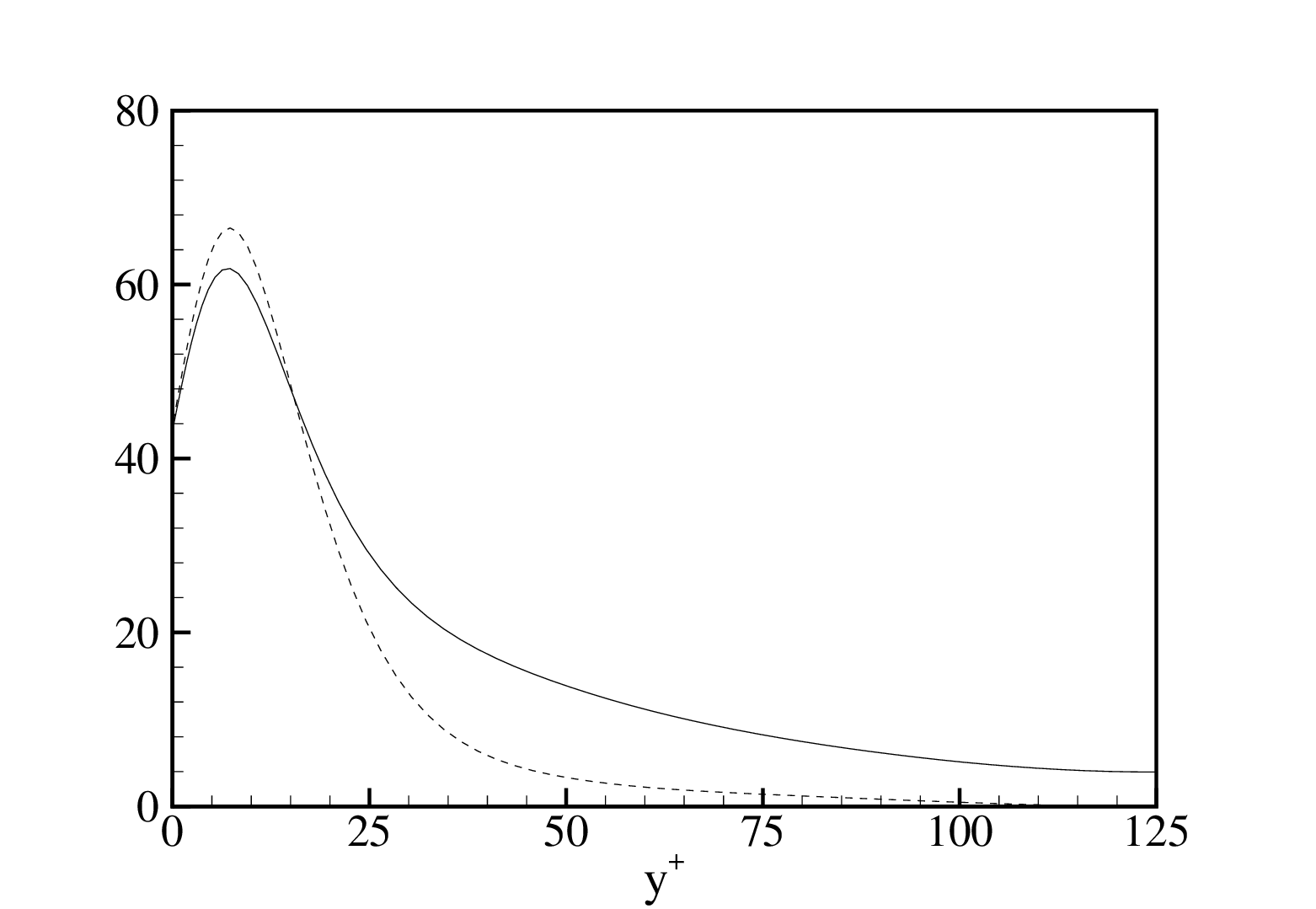,width=7.0cm} 
          \hspace{-0.4cm}
          \epsfig{figure=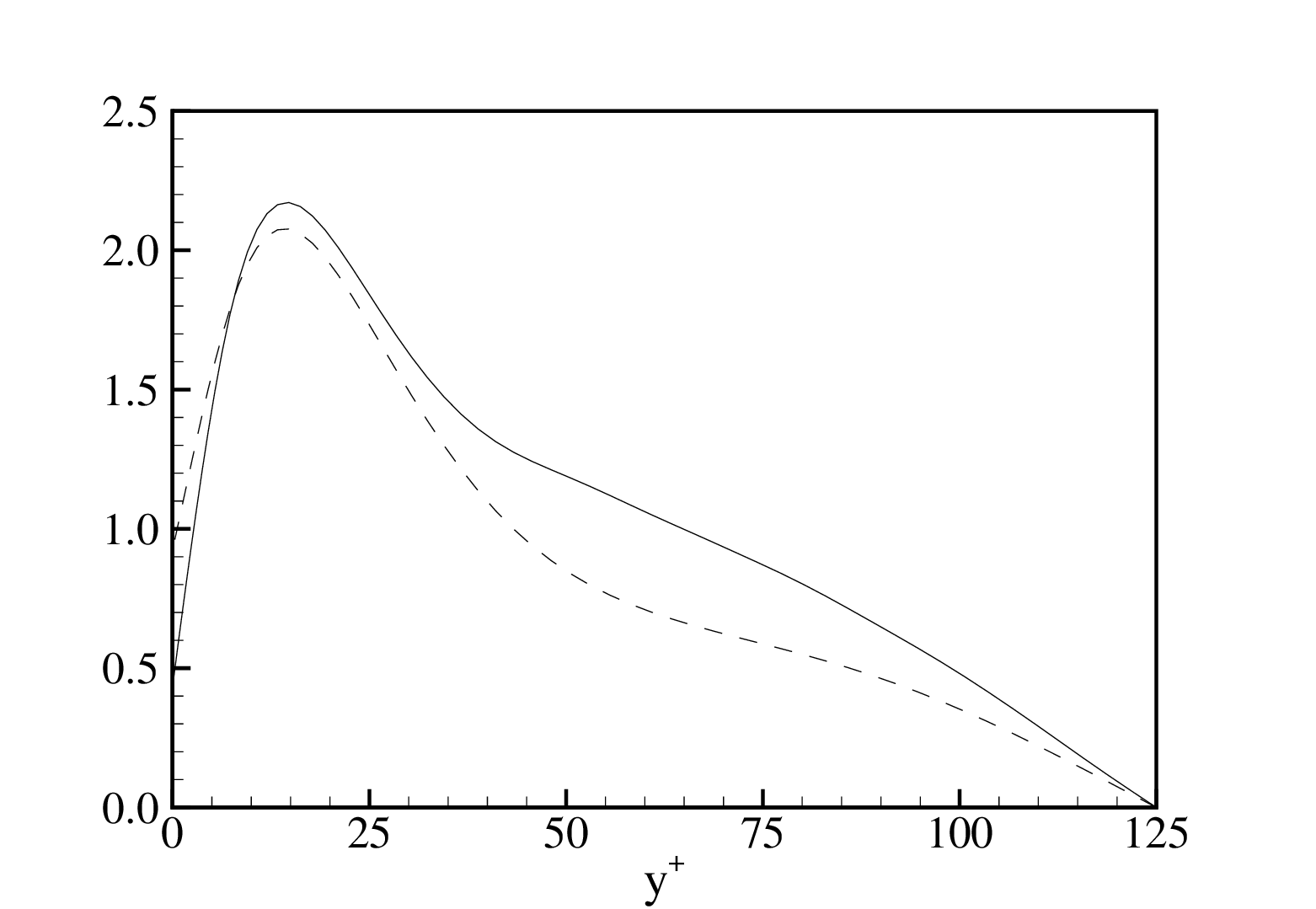,width=7.0cm} 
           } 
\caption{Left panel: quality of the 1-point closure for equation (\ref{xx}) as obtained by direct
numerical simulation for a turbulent channel flow, see De Angelis et. al. (2003).
The continuous line refers to the term $2 a_x R_{xy}S$ with $a_x=1$ while the
dashed line to the term $\langle P \rangle \langle R_{xx}\rangle /\tau$. 
The the two terms are of the same order of magnitude
within the buffer region.
Right panel: comparison of the 1-point closure approximation for
equation (\ref{xy}). The continuous line refers to the term
$a_yR_{yy} S$ with $a_y=0.45$, while the dashed line to the term $ \langle P \rangle
\langle R_{xy}\rangle /\tau $. The the two terms are of the same
order of magnitude within the buffer region. }
\label{eqxx} 
\efi

In Figs. (\ref{eqxx})  we show the comparison of the 1-point closure approximation, using numerical simulations
with $a_x= 1$ and $a_y = 0.45$.  The overall comparison is
reasonably good within the buffer layer, i.e. in the region where
the polymers enhance the mean flow with respect to the Newtonian
case. We can thus safely proceed and learn from Eq. (\ref{xx})
that in 1-point closure we can write the third term on the LHS of
Eq. (\ref{balmean}) as
\begin{equation}
\frac{\nu_p}{\tau} \langle P R_{xy}\rangle (y)S = \frac{\nu_p}{2\tau^2} 
\langle P\rangle^2 \langle R_{xx}\rangle \ ;
\end{equation}
as such, it is one of the three terms appearing in the trace in
Eq. (\ref{epsp}). As this term is already exhausted in the balance
for the mean energy $V^2$, this leaves for the turbulent energy
balance equation the other two terms. We stress here that this
conclusion rests entirely on the numerical value $a_x= 1$. We
propose that this is essentially exact, even though at this stage
we present it on the basis of numerical comparisons. In future
work we will derive this important fact from first principles.

We can therefore write now the balance equation for the turbulent
energy. The rate of production is simply $W(y)S(y)$. This
production of energy is balanced by the energy dissipation
$\nu\langle  |\nabla u|^2\rangle$ and the two terms in the
energy transferred to the polymer (\ref{epsp}). The energy
dissipation is estimated differently in the viscous layer near the
wall and in the bulk of the turbulent flow, namely 
\begin{equation}
a\nu\frac{K}{y^2}+ b\frac{K^{3/2}}{y}+\frac{A^2 \nu_p}{2
\tau^2} \langle P\rangle^2(\langle R_{yy}\rangle +\langle
R_{zz}\rangle) = WS \ , \label{2}
\end{equation}
where the first term reflects the smoothness of the field in the viscous
layer, and the second term is a standard Kolmogorov-type estimate
of the energy flux as the typical energy at distance $y$ from the
wall over the typical eddie turn-over time $y/\sqrt{K(y)}$. These
estimates are identical to the Newtonian case. The new (and
somewhat difficult) term is the third term on the LHS, where $A$
is a constant to be computed.

To evaluate this last term, we refer again to our DNS results to assess the relative
importance of $\langle R_{yy}\rangle$ and $\langle R_{zz}\rangle$:
these objects are very close to each other throughout the region
of concern in the channel. We therefore can keep only one of them
by redefining consequently the constant $A$:
\begin{equation}
a^2\nu\frac{K}{y^2}+ b\frac{K^{3/2}}{y}+\frac{A^2 \nu_p}{2
\tau^2} \langle P\rangle^2\langle R_{yy}\rangle = WS \ .
\label{balEvis}
\end{equation}

\subsection{Reminder: the {\VK} law}
At this point it is useful to remind to the reader the derivation of
the {\VK} log law of the wall in the Newtonian case, see 
\cite{00Pope}. The terms including the conformation tensor in Eqs.
(\ref{1}) and (\ref{balEvis}) are not present in this case. Outside
the boundary layer the viscous terms proportional to $\nu$ are
neglected, and we therefore get from Eq. (\ref{1}) the result that
$W$ is constant, $W\approx p'L$, and from Eq.(\ref{balEvis}) that
$bK^{3/2}/y\sim WS$. Now one introduces the experimental knowledge
that $K$ and $W$ are proportional to each other outside the viscous
boundary layer:
\begin{equation}
W =c_ N^2 K \ . \label{WK}
\end{equation}
This equation can be presented as a rigorous inequality $W\le K$
(i.e. $c_N$ is at most unity), and it can be justified
phenomenologically as a result of the scale invariance. We will
take it, and its visco-elastic counterpart
\begin{equation}
W =c_V^2 K \ , \label{WKvis}
\end{equation}
as given experimental inputs. Using (\ref{WK}) in the relations obtained
for the Newtonian case, we find immediately $S\propto
\sqrt{p'L}/y$ which integrates to the log-law of the wall $V=\log
y/\kappa_K +B$ where the {\VK}  constant $\kappa_K$
and the intercept $B$ are phenomenological constants.
\subsection{Derivation of the MDR}
We anticipate that in the viscoelastic case the dominant terms in
the balance equations (\ref{1}) and (\ref{balEvis}) will be the
terms including the polymer conformation tensor. To see that this
is indeed the case, we analyze the $yy$ component of Eq.
(\ref{3}). It reads
    \begin{eqnarray}\label{yycom}
\frac{\partial}{\partial y} \langle u_y R_{yy}\rangle
=-\frac{1}{\tau}\langle PR_{yy}\rangle +2 \langle R_{yk}\partial_k
u_y\rangle \ .
    \end{eqnarray}%
In 1-point closure this equation gives the absurd result $\langle
PR_{yy}\rangle =0$ which means that in
this case we must proceed to the next order and estimate

\begin{eqnarray}\label{Ryk}
\la R_{yk} \partial_k u_y \ra 
 \approx \la
R_{yy}\ra \left[K(y)/y^2\right]^{1/2} \approx c_V \la
R_{yy}\ra \left[W(y)/y^2\right]^{1/2} \ .
\end{eqnarray}%
Using this this estimate we can proceed to conclude from Eq.
(\ref{yycom}) that
\begin{equation}
\frac{1}{\tau}\la P \ra \la R_{yy}\ra = g \la R_{yy}\ra
W^{1/2}(y)/y\,, \label{Ryy}
\end{equation}
where $g$ is an empirical constant. Because equation (\ref{Ryy})
is crucial for the evaluation of the momentum flux in the buffer
layer, we show in Fig. \ref{Wtheory} (left panel) the quality of the closure
approximation as obtained by numerical simulations. As one can
see, the quality of the approximation is rather good within the
buffer layer.

\bfi[t!] 
          \centerline{ 
          \epsfig{figure=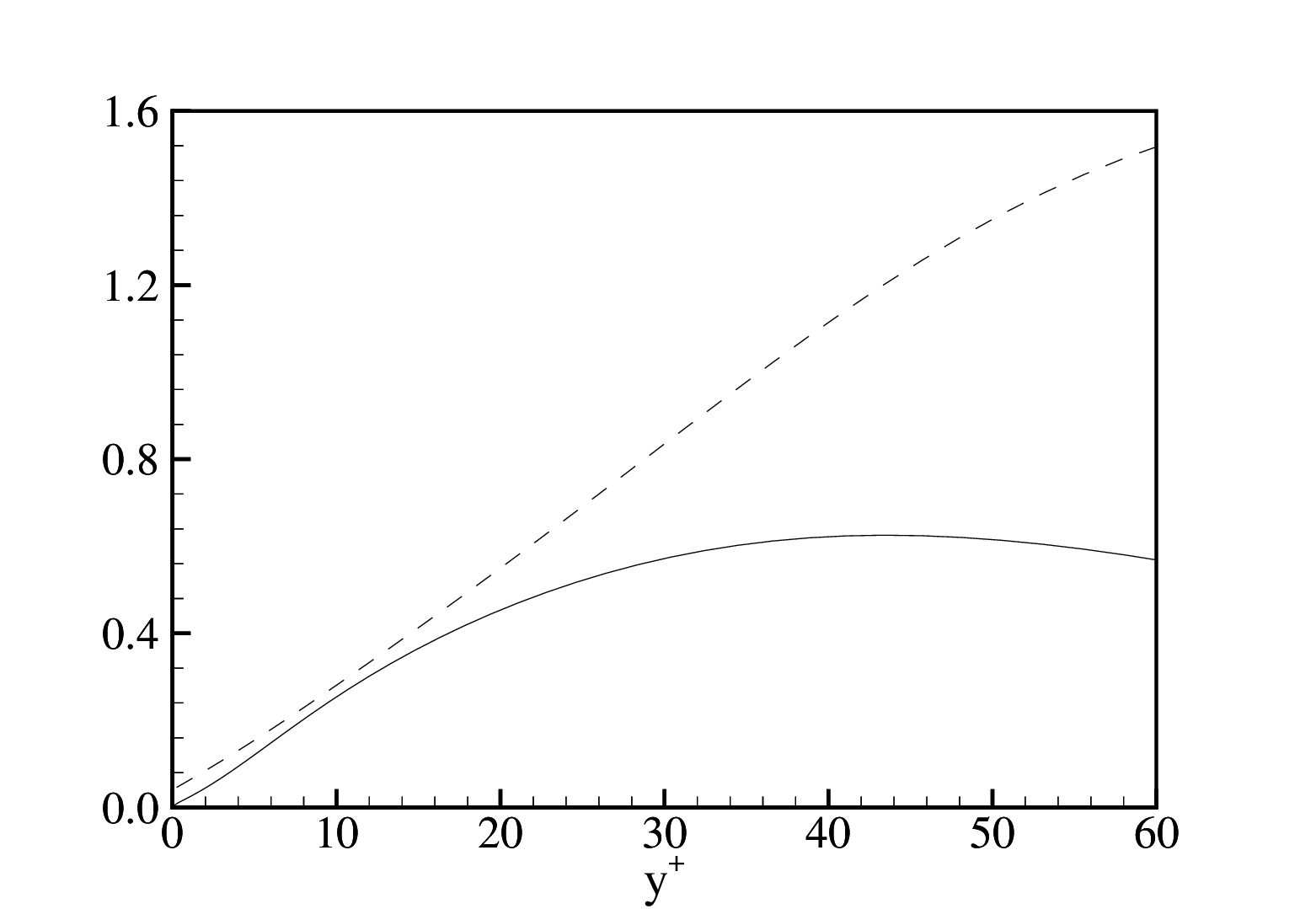,width=7.0cm} 
          \hspace{-0.4cm}
          \epsfig{figure=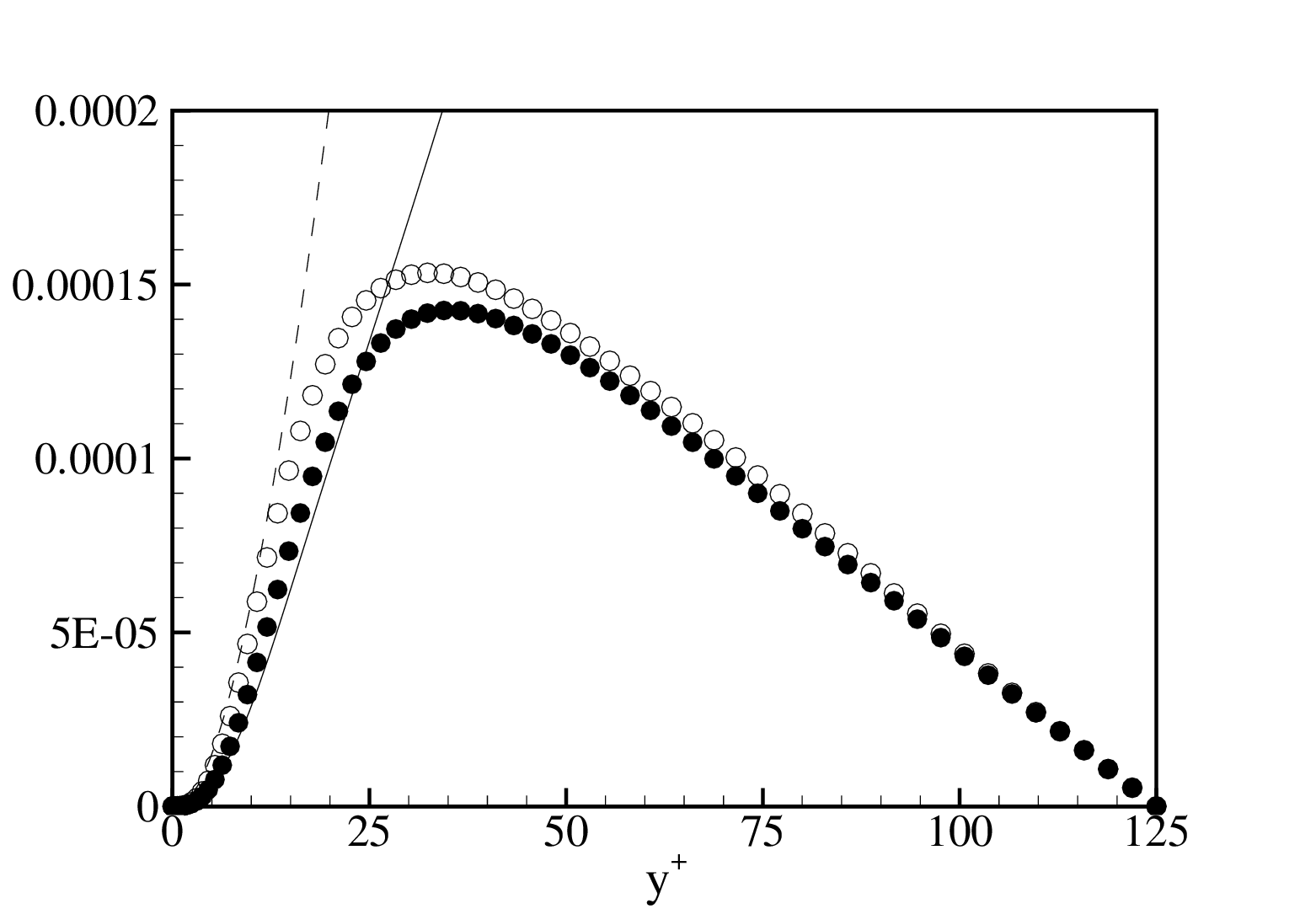,width=7.0cm} 
           } 
\caption{Left panel:Comparison between the term $\langle R_{yy}\rangle
\langle P \rangle/\tau$ and $\langle R_{yy}\rangle W^{1/2}/y$ in
the numerical simulations. Right panel:Comparison of $W$ as obtained by direct numerical
simulations against the theoretical estimate $y^2\langle P
\rangle^2/\tau^2$ for two different values of $\tau$, namely
$\tau=12.5$ and $\tau=37.5$. The estimate of $g$ for the two cases gives
$g=1.22$ for $\tau=12.5$ and $g=1.1$ for $\tau=37.5$. Thus, although the value of
$\tau$ changes by a factor $9$, the combination of the Peterlin function with the $\tau^{-2}$ 
term gives the same overall constant within $10$ {\it per cent} error, showing the excellent 
quality of the prediction (\ref{4})
\label{Wtheory} }
\efi

This allows us to offer a final result for $W(y)$ in which the
polymer conformation tensor disappears altogether:
\begin{equation}
W(y) \sim \frac{\la P\ra^2 y^2}{g^2 \tau^2 } \ .
\label{4}
\end{equation}
This is an important result. On the one hand it shows how the
Reynolds stress is reduced in the elastic layer, providing a
direct mechanism to drag reduction. The Reynolds stress is the
momentum transfer to the wall, and reducing it results in an
increase of the mean momentum, in order to balance the constant
rate of momentum production $p'L$. On the other hand, we see that
this quantity is becoming of $O(y^2)$, justifying the neglect of
Reynolds stresses in the simplified theory of \cite{03LPPT}. In
Fig. \ref{Wtheory} (right panel) we compare the prediction of equation (\ref{4})
against the direct numerical simulations. The quality of the
results are extremely good, showing that the right physical
behavior has been captured by employing the closure
approximations.

Neglecting then the first two terms in eq. (\ref{1}), without the term $p' y$, and eq
(\ref{balEvis}), and using the result (\ref{xy}) for the
conformation tensor we end up with the estimate for the shear
$S(y)$ in the form
\begin{equation}
S \sim \frac{A g}{\sqrt{2 a_y}} \frac{ \sqrt{p'L}}{ y} \ .\label{5}
\end{equation}
Let us note that (\ref{5}) is independent of $\tau$ and it is
consistent with the findings of \cite{03LPPT}, i.e. it predicts a
new logarithmic law of the wall which is the MDR asymptote. We
note that the dissipative terms did not play any role in the
viscoelastic derivation; the polymer terms took over, providing
precisely the same role of the effective viscosity introduced in
the simplified theory of \cite{03LPPT}.

In order to compute the value of $A$, we invoke the estimate of
$a$ in equation (\ref{balEvis}) as obtained in \cite{03LPPT}, i.e.
$a = y_v^+ c_N$ where  $y_v^+$ is the thickness of the viscous
sub-layer, for the Newtonian fluid, expressed in the Prandtl units
$y^+ \equiv y\sqrt{p'L}/\nu$. Next, we compute the value of $y_F$ 
for which the MDR mean flow shear is equal to the viscous mean
flow shear:
$$
 \frac{A g}{\sqrt{2a_y}} \frac{ \sqrt{p'L}}{ y_F} = \frac{p'L}{\nu}\,,
$$
which gives, in Prandtl units,
\begin{equation}
y^+_F = \frac{A g}{\sqrt{2a_y}} \ . \label{yf}
\end{equation}
For $y=y_F$ we expect that the energy dissipation due to
viscous effect $a \nu K/y_F^2$ is of the order of the energy
dissipation due to polymers stretching, i.e.
$$
a^2 \nu \frac{K}{y_F^2} \ \sim \ \frac{A^2 \nu_p}{2 \tau^2}
\langle P\rangle^2(\langle R_{yy}\rangle\ .
$$
Finally, by using the MDR estimate (\ref{5}) and equations
(\ref{xy}),(\ref{4}, and (\ref{yf}), we can compute the value of $A$.
It turns out that for the MDR slope $A g / \sqrt{2a_y} \sim y_v^+
(c_N/c_V)$,  in agreement with the previous findings of
\cite{03LPPT}. 
\section{Saturation of drag reduction and cross over from the MDR
to the Newtonian plug}
We expect a cross over from the MDR asymptote back to the
Newtonian plug when the basic assumptions on the relative
importance of the various terms in the balance equations lose
their validity. Experimentally one sees that mean velocity profile
follows the MDR up to some point $y^+_V$ after which it
crosses over back to a logarithmic profile with the same slope as
the Newtonian flow.  To measure the quality of drag reduction one
can introduce a dimensionless drag reduction parameter
\begin{equation}\label{Q}
 Q \equiv \frac{y^+_{V}}{y^+_{N}} -1\ .
\end{equation}
The Newtonian flow is then a limiting case of the viscoelastic
flow corresponding to $Q=0$. There can be a number of reasons for
the saturation of the drag reduction effect. One reason that can
be important for large Reynolds numbers is that the
concentration of the polymer is too small to provide the necessary
increase in the effective viscosity as a function of the $y^+$.
This mechanism for the cross over back to the Newtonian plug was
discussed in full detail in \cite{04BDLPT}.  The main result of
\cite{04BDLPT} is that the drag reduction parameter $Q$ is given
by
\begin{equation}\label{sigma1}
  Q = \alpha^3c_p N_p ^3 \;\;\; c_p small \; , Re \; large
\ ,
\end{equation} 
where $c_p$ is the number density
(concentration in number of molecules per unit volume) of the
polymer and $N_p$ is the degree of polymerization (the number
of monomers per molecule). The parameter $\alpha$ is the linear
scale of the monomer. This prediction was tested in \cite{04BDLPT}
by comparing with experiments with DNA as the drag reducing agent,
with excellent agreement between theory and experiments.

Here we address another mechanism for the saturation of drag
reduction.  We expect a cross over from the MDR asymptote back to the
Newtonian plug when the basic assumptions on the relative
importance of the various terms in the balance equations lose
their validity, i.e. when (i) the turbulent momentum flux $W$ becomes comparable
to the total momentum flux $p'L$, or when (ii) the turbulent energy flux
$bK^{3/2}/y$ becomes of the same order as the turbulent energy production $WS$.
Using the estimates (\ref{4}) and (\ref{5}), one  finds that both these conditions
give the same cross-over point
\begin{equation}\label{co1}
  y_V \simeq \frac{\tau\sqrt{p'L}}{\la P\ra g}
\ .\end{equation}
Note that $\tilde\tau(y)\equiv\tau/\la P(y)\ra$ is the effective non-linear
polymer relaxation time. Therefore Eq. (\ref{co1}) can be also rewritten
as
\begin{equation}\label{co2}
  S(y_V)\tilde\tau(y_V) \simeq 1
\ .\end{equation}
In writing this equation we use the fact that the cross over point belongs also
to the edge of the Newtonian plug where $S(y)\approx \sqrt{p'L}/y$ and that $g\sim 1$. The LHS of this
equation is simply the \textit{local Weissenberg number} (product of local mean shear and local effective
polymer relaxation time). Thus, \textit{ the cross-over to the Newtonian plug occurs at the point, where 
the local Weissenberg number decreases to $\sim1$}. We expect that this result
remains valid for any model of the elastic polymers, not only for the
FENE-P model considered here.

To understand how the cross-over point $y_V$ depends on the polymer
concentration and other parameters, we need to estimate mean value of the
Peterlin function $\la P\ra$.
Let us estimate the value of $\langle P \rangle$ as
$$
\langle P \rangle = \frac{1}{1-\alpha \langle R\rangle }\,,
$$
where $\langle R\rangle = \langle R_{xx}+R_{yy}+R_{zz}\rangle \sim
\langle R_{xx}+2R_{yy}\rangle $ and $\alpha \sim 1/\rho_m^2$ (for
simplicity we disregard $\rho_0$). It follows from Eqs.~(\ref{xx})-(\ref{xy}),
that
$$
  \la R_{xx}\ra \simeq (S\tilde\tau)^2 \la R_{yy}\ra
\,,$$
and at the cross-over point (\ref{co2})
$$
  \la R_{xx}\ra \simeq \la R_{yy}\ra
\,,\quad
  \la R\ra \simeq \la R_{yy}\ra
\ .$$
The dependence of $\la R_{yy}\ra$ on $y$ in the MDR region follows from
Eqs.~(\ref{balEvis}), (\ref{4}), and (\ref{5}):
$$
  \la R_{yy}\ra \simeq \frac{\tilde\tau^2WS}{\nu_p} \simeq
    \frac{y\sqrt{p'L}}{\nu_p}
\ .$$
Then at the cross-over point $y=y_V$:
$$
 \la P\ra \simeq \frac1{1-\alpha y_V\sqrt{p'L}\big/\nu_p}
\ .$$
Substituting this estimation into (\ref{co1}) gives the final result
\begin{equation}\label{co3}
 y_V = \frac{C\tau\sqrt{p'L}}{1+\alpha p'L\tau\big/\nu_p}
\ .\end{equation}
Here $C$ is constant of the order of unity.  Finally, introducing the dimensionless concentration of
the polymers
\begin{equation}
  \tilde c_p \equiv \frac{\nu_p}{\alpha\,\nu_0}
\,,\end{equation}
we write the denominator in Eq.~(\ref{co3}) as
$$
 1 +\frac{\alpha\,p'\,L\,\tau}{\nu_p} = 
   1 +\frac{1}{\tilde c_p}\frac{p'\,L\,\tau}{\nu_0} = 
   1 +\frac{{\rm We}}{\tilde c_p} \ .
$$
Here
\begin{equation}
  {\rm We} \equiv \frac{p'\,L\,\tau}{\nu_0}
\end{equation}
is the (global) Weissenberg number. The final result for the dimensionless cross-over point
$y_V^+\equiv y_V\sqrt{p'L}\big/\nu_0$ assumes a very simple form:
\begin{equation}\label{co4}
  y_V^+ = \frac{C{\rm We}}{1+{\rm We}/\tilde c_p}
\ . \label{crossfi}
\end{equation}

This prediction can be put to direct test when $\tilde c_p$ is very large,
or equivalently in the Oldroyd B model where $P\equiv 1$ \cite{03BCHP}. Indeed, in numerical simulations
when the Weissenberg number was changed systematically, cf. \cite{01YKTM}, one observes the cross over
to depend on We in a manner consistent with Eq. (\ref{crossfi}). The other limit when 
$\tilde c_p$ is very small is in agreement with the linear dependence on $\tilde c_p$
predicted in Eq. (\ref{sigma1}).

We can thus reach conclusions about the saturation of drag
reduction in various limits of the experimental conditions, in
agreement with experiments and simulations. 

\section{Conclusions}

In this paper we showed how to understand some of the most common
effects of dilute polymers solutions in wall bounded turbulent
flows. Our starting point is the FENE-P model which has been shown
to provide good qualitative and quantitative agreement with
experimental findings. Within the context of the FENE-P model, we
were able to derive in a controlled way the universal MDR profile
that was first obtained theoretically in \cite{03LPPT}. One of our
main predictions is provided by Eq. (\ref{4}) which tells us how
the Reynold stress  behaves as  a function of the Peterlin
function and the polymer relaxation time $\tau$. The predictions
are in very good agreement with the results of DNS. In particular,
Eq. (\ref{4}) shows that the Reynolds stress is proportional to
$\langle P \rangle^2\tau^{-2}$, i.e. it becomes small for long
polymers chains ($\tau$ large) and large concentration. The
velocity profile comes out independent on the concentration up to
a point, $y^+_V$, given by Eq. (\ref{crossfi}), where the {\VK}  
law is restored.  In general our theoretical approach rationalizes and
predicts a large number of qualitative and quantitative aspects of
drag reduction as seen in laboratory experiments and in DNS of the
FENE-P model.

\acknowledgments \vskip 0.5 cm This work was supported in part by
the European Commission under a TMR grant, the US-Israel
Binational Science Foundation, and the Minerva Foundation, Munich,
Germany.


\begin{thebibliography}{}

\bibitem[Lumley (1969)]{69Lu}
{\sc Lumley J.L.} 1969 {\it Ann. Rev. Fluid Mech.} {\bf 1}, 367.

\bibitem[Sreenivasan \& White (2000)]{00SW}
{\sc Sreenivasan K.R. White C.M.} 2000 {\it J. Fluid Mech. } {\bf 409}, 149.

\bibitem[Virk (1975)]{75Virk}
{\sc Virk P.S.} 1969 {\it J. AIChE} {\bf 21}, 625.


\bibitem[De Angelis et. al. (2003)]{03ACLPP}
{\sc De Angelis E., C.M. Casciola, V.S. L'vov, R. Piva and I. Procaccia} 2003
{\it Phys. Rev. E.} {\bf 67}, 056312 .

\bibitem[Warholic et. al. (1999)]{99WMH}
{\sc Warholic M.D., H.Massah, T.J. Hanratty} 1999 {\it Experiments in Fluids}, {\bf
27}, 461.

\bibitem[Rollin \& Seyer (1972)]{72RS}
{\sc Rollin A.,F.A. Seyer} 1972 {\it Can. J. Chem. Eng},{\bf
50}, 714-718.

\bibitem[L'vov et. al. (2003)]{03LPPT}
{\sc L'vov V.S., A. Pomyalov, I. Procaccia, V. Tiberkevich} 2003 {\it Phys. Rev. Lett.}
submitted, also: nlin.CD/0307034.

\bibitem[Rudd (1969)]{69Rud}
{\sc Rudd M.J.}, 1969 {\it Nature}, {\bf 224}, 587.

\bibitem[De Angelis et. al. (2004)]{04DCLPPT}
{\sc De Angelis E., C. Casciola, V. S. L'vov, A. Pomyalov, I. Procaccia, V. Tiberkevich}
2004    Drag Reduction by a linear viscosity profile, {\it Phys. Rev. E}, submitted, also: nlin.CD/0401005

\bibitem[Bird et. al. (1987)]{87BCAH}
{\sc Bird R.B., C.F. Curtiss, R.C. Armstrong, O. Hassager}, 1987 {\it
Dynamics of Polymeric Fluids} {\bf Vol.2 } (Wiley, NY).

\bibitem[Beris \& Edwards (1994)]{94BE}
{\sc Beris A.N., B.J. Edwards}1994, {\it Thermodynamics of
Flowing Systems with Internal Microstructure} (Oxford University Press, NY).

\bibitem[Pope (2000)]{00Pope}
{\sc Pope S.C.}, 2000 {\it Turbulent Flows} (Cambridge).

\bibitem[Benzi et. al. (2004)]{04BDLPT}
{\sc Benzi R., V. S. L'vov, I. Procaccia V. Tiberkevich }, 2004 {\it Saturation of Turbulent Drag Reduction in
    Dilute Polymer Solutions} Phys. Rev. Lett in press, Also:nlin.CD/0402027

\bibitem[Benzi et. al. (2004)]{03BCHP} 
{\sc Benzi R., E. Ching, N. Horesh, I. Procaccia}, 2004, Theory of
concentration
dependence in drag reduction by polymers and of the MDR asymptote, {\it Phys.
Rev. Lett. Feb} {\bf 20 }.

\bibitem[Yu et. al. (2001)]{01YKTM}
{\sc Yu B., Y. Kawaguchi, S Takagi, Y. Matsumoto} 2001, The 7th symposium on smart
control of turbulence, University of Tokyo.


\end{thebibliography}
\end{document}